\documentclass[preprint2]{aastex}

\newcommand{\ea}{et al.}

\newcommand{\kms}{\>{\rm km}\,{\rm s}^{-1}}

\newcommand{\kpc}{\>{\rm kpc}}

\newcommand{\bfi}{\begin{figure}[htb]} 
\newcommand{\bpfi}{\begin{figure}[p]}

\newcommand{\halpha}{$\rm H\alpha$}
\newcommand{\hbeta}{$\rm H\beta$}
\newcommand{\as}{ ^{\prime\prime}}
\newcommand{\am}{^{\prime}}

\newcommand{\DensePak}{{\texttt{DensePak}}}
\newcommand{\nii}{\mbox{[N\,{\sc ii}]}}
\newcommand{\sii}{\mbox{[S\,{\sc ii}]}}

\def\deg{\hbox{$^\circ$}}
\def\hii{\relax \ifmmode {\rm H\,{\sc ii}}\else H\,{\sc ii}\fi}
\def\hi{\relax \ifmmode {\rm H\,{\sc i}}\else H\,{\sc i}\fi}

\def\fdg{\hbox{$.\!\!^\circ$}}

\def\farcm{\hbox{$.\mkern-4mu^\prime$}}
\def\farcs{\hbox{$.\!\!^{\prime\prime}$}}
\def\degd#1.#2{ #1\fdg#2 }            
\def\mind#1.#2{ #1\farcm#2 }               
\def\secd#1.#2{ #1\farcs#2 }               

\def\aj{AJ}                   
\def\apj{ApJ}                 
\def\apjl{ApJ}                
\def\apjs{ApJS}               
\def\apss{Ap\&SS}             
\def\aap{A\&A}                
\def\mnras{MNRAS}             
\def\nat{Nature}              

\slugcomment{initial submission}

\shorttitle{}
\shortauthors{Mazzuca, et al.}

\begin{document}
\title{Nuclear Rings in Galaxies---A Kinematic Perspective}

\author{Lisa. M. Mazzuca}
\affil{NASA Goddard Space Flight Center, Mission Validation and Operations Branch (Code 584), Greenbelt, MD 20771, USA}

\author{Robert A. Swaters}
\affil{Department of Astronomy, University of Maryland, College Park MD 20742, USA}

\author{Johan H. Knapen}
\affil{Instituto de Astrof\'{i}sica de Canarias, E-38200 La Laguna, Spain and}
\affil{Departamento de Astrof\'\i sica, Universidad de La Laguna, E-38205 La Laguna,Spain}

\author{Sylvain Veilleux}
\affil{Department of Astronomy, University of Maryland, College Park MD 20742, USA}

\begin{abstract}
We combine DensePak integral field unit and TAURUS Fabry-Perot observations of 13 nuclear rings to show an interconnection between the kinematic properties of the rings and their resonant origin. The nuclear rings have regular and symmetric kinematics, and lack strong non-circular motions. This symmetry, coupled with a direct relationship between the position angles and ellipticities of the rings and those of their host galaxies, indicate the rings are in the same plane as the disc and are circular. From the rotation curves derived, we have estimated the compactness ($v^2/r$) up to the turnover radius, which is where the nuclear rings reside. We find that there is evidence of a correlation between compactness and ring width and size. Radially wide rings are less compact, and thus have lower mass concentration. The compactness increases as the ring width decreases. We also find that the nuclear ring size is dependent on the bar strength, with weaker bars allowing rings of any size to form. 
\end{abstract}


\section{Introduction}
 
Nuclear rings in the central kiloparsecs of galaxies are usually observed as conglomerations of young massive stars in distinct compact groupings \citep{BC96, JK05, C10}. The star forming rings are believed to form as a result of radial gas inflow towards the central region, which stagnates near dynamical resonances. In barred galaxies the inflow originates in the gravitational torques set up by the bar (e.g., \cite{A92,  HS94, HS96, P95, JK95, BC96, R97, B02}). Some 20\% of nuclear rings occur in non-barred galaxies \citep{JK05, C10} and in those cases the inflow and resonances may well be related to the presence of weak ovals,  past interactions or mergers, or even strong spiral arms. A resonant origin of all nuclear rings is most plausible.
 
The radial location of a nuclear ring can be described by the position of the inner Lindblad resonance(s) (ILRs) \citep{S80, Sh89, BC93, FB93, HS94, JK95, BC96}, which is typically at $\sim$1 kpc from the galaxy center although smaller and larger nuclear rings exist \citep{BC93, C10}. In a non-axisymmetric potential, the nuclear ring forms near the ILR(s), where the inflowing molecular gas slows down, accumulates, and initiates star formation. For barred spiral galaxies, the ring can be formed inside the large scale galactic spiral structure in the vicinity of the inner Lindblad resonances (ILRs), which occur at those radii where the bar pattern speed $\Omega_{b}$ equals ($\Omega - \kappa/2$), with $\Omega$ the angular velocity and kappa the radial epicyclic frequency. This approximation is not valid for strong bars where a non-linear analysis must be performed, but in any case serves as a useful indication of how the location of the ring is related to the underlying galaxy dynamics \citep{JK95}.

Only a few in-depth observational studies of the kinematics of nuclear rings at high angular resolution exist, but these have revealed some consistent kinematic characteristics of nuclear rings. \cite{B96} showed that non-circular motions near the intersection of the ring and bar dust lanes correspond to radially inflowing molecular gas that is apparently feeding the ring in NGC~4314. \cite{RD98} concluded for NGC~1530 the existence of weak velocity transitions at the ends of the bar which are inversely correlated with strong star formation in the same areas, including the nuclear ring location. \cite{R97} further note an increase in the residual velocity along an axis perpendicular to the bar. \cite{JK00} (see also \cite{JK95}) and \cite{Z04} used CO and \halpha\ emission in the central region of NGC~4321 and NGC~1530, respectively, to reveal gas streaming inward towards the ring along the bar dust lanes. \cite{J02} studied the fueling mechanisms of the ring in NGC~5248 to deduce that large star forming clusters in the ring have been triggered by a bar-driven spiral density
  wave. More recently, \cite{A06} follow up on the ring in NGC~4321 to show relatively low \hbeta\ gas dispersion within the ring, which can be an indicator of active star formation and cool gas inflow from which the massive stars inherited the low velocity dispersion.

The existing kinematic studies provide insight into a small number of nuclear rings, but fail to reach more general conclusions applicable to the overall population. Apart from the differing characteristics of the data used in these studies, this is due mostly to the intrinsically small angular size of nuclear rings, and the dusty environments in which they exist, both of which make rings non-trivial to observe with most common instruments yielding kinematic data. In this paper we endeavor to  start rectifying this situation by presenting a more homogeneous data set on thirteen nuclear rings.

The nuclear rings studied spectroscopically in the current paper were all part of the sample of \cite{M08} (hereafter known as M08), who provided photometric insight into the strong star forming nature of nuclear rings. From their \halpha\ observational study of 22 nuclear rings, they find that many of the rings exhibit a well-defined age distribution pattern. M08 conclude that this occurrence can be a result of the combination of bar-induced dynamics and gravitational instabilities which are occurring in the proximity of the rings. These results add credence to the idea that the mass inflow driven by the bar along the dust lanes to the ring contact points is a key requirement for the fueling of active (i.e., star forming) nuclear rings (and might also yield a possible signature in the form of non-circular motions). Further evidence for this idea from detailed observations of a handful of galaxies has been presented by, e.g., \cite{A06, B08}, and \cite{JK10}. 

We now use kinematic information within and just outside the nuclear rings in 13 galaxies to investigate the possible relations between the optical morphology of the rings, the parameters derived from the rotation curve, and their resonant nature. We begin by presenting an overview of the observations and morphological properties of the \halpha\ sample in Section~2. We discuss the basic data reduction method, which includes the construction of the velocity field maps, in Section~3. In Section~4 we discuss the kinematic parameters associated with the ring rotation curves, and present results pertaining to non-circular motions in and around the nuclear rings, including velocity residuals. Our detailed analysis of the results is discussed in Section~5, with concluding remarks in Section~6.

\section{Sample and Observations} 

The galaxies studied here (see Table~\ref{tab:phot}) form a subset of the sample presented in the imaging survey of M08 (which was based on $B$, $I$, and \halpha\ imaging). The data in M08 were obtained to characterize the morphology of the nuclear rings, and to analyze the age distribution of the individual stellar clusters forming each ring using the \halpha\ equivalent width. Results from this photometrically based survey reveal that the rings contain very young (1 Myr to 10 Myrs) hot massive stars, with some of the youngest clusters located in proximity to one or both of the intersection points of the bar dust lanes and the outer edge of the ring.  The ring ellipticity and position angle approximately match that of the respective galactic disk, indicating that the rings are in the same plane as the disk and are circular (M08; see also \cite{C10}).
  
\begin{table*}
\begin{center}
\footnotesize{
\caption[]{Morphological and photometric parameters of \DensePak\ sample}
\label{tab:phot}
\begin{tabular}{lccccccccccccc}
\hline \hline
NGC & Morph & \multicolumn{2}{c}{Offset} & $i$ & ${\rm \phi_{r}}$ & ${\rm \phi_{d}}$ &${\rm \phi_{b}}$ & ${\rm \epsilon_{r}}$ & ${\rm \epsilon_{d}}$& ${\rm Radius}$ & ${\rm Radius}$  & {\rm Run}\\
& Type & $x$ & $y$ & & & & & & &  & &Date\\
 & & ($\as$) & ($\as$) &($\deg$) & ($\deg$)& ($\deg$) & ($\deg$) & &&($\as$) & (kpc) & \\
(1) & (2) & \multicolumn{2}{c}{(3)} & (4) & (5) & (6) & (7) & (8) & (9) & (10) & (11) & (12)\\
\hline
~473  &  SAB(r)0/a & $-$4.4 & +7.2 &49 &  154 & 153 & 164 & 0.37&0.37&12.2 $\times$ 6.9 & 1.7 $\times$ 1.0 &  Dec03\\
1300 &  SB(rs)b & n/a & n/a& 40 & 135 & 106 & 102 & 0.25 & 0.34 & 4.1 x 3.1 & 0.3 x 0.2 &  Sep98\\
1343  &  SAB(s)b  & $-$4.4 & +7.8 & 41 &  60  & 80 & 82& 0.25& 0.37&8.8 $\times$ 6.6 & 1.2 $\times$ 0.9 &  Dec03 \\
1530  &  SB(rs)b  & $-$1.6 & +11.3 & 48 &  25  &8& 122 & 0.35&0.29 & 6.8 $\times$ 4.9 & 1.2 $\times$ 0.8 & Dec03 \\
2903  &  SAB(rs)bc   & $-$6.8 & +8.3 & 65 &  8   & 17 & 24&  0.4& 0.52 &1.7 $\times$ 1.0 & 0.06 $\times$ 0.04  & Dec03 \\
3351  &  SB(r)b   & $-$4.2 & +6.1 & 34 &  20  & 13 &  112&  0.3 & 0.32&2.7 $\times$ 1.9 & 0.15 $\times$ 0.11  & Dec03 \\
4303  &  SAB(rs)bc & $-$3.3 & +7.5 & 18 &  88  &$-$ & 10 &0.14&0.11&   3.3 $\times$ 2.8 & 0.2 $\times$ 0.2 &  Apr04 \\
4314  &  SB(rs)a   & $-$3.0 & +6.3 & 26 &  135 &$-$& 135 &0.1&0.11&   6.6 $\times$ 5.9 & 0.3 $\times$ 0.3 &  Apr04\\
4321 &  SAB(s)bc & n/a & n/a  & 30 & 170 & 30 & 153 & 0.12 & 0.15& 8.8 x 7.0 & 0.7 x 0.6 &  May95\\
5248  &  SAB(rs)bc  & $-$3.3 & +6.5 & 46 &  115 & 110 &137&0.3&0.28& 6.6 $\times$ 4.6 & 0.7 $\times$ 0.5 & Apr04 \\
5953  &  SAa       & $-$6.0  & +6.6 & 26 &  172 & 169 & no bar&0.1&0.17& 6.1 $\times$ 5.5 & 1.0 $\times$ 0.9 & Apr04 \\  
6951 &  SAB(rs)bc & n/a  & n/a  & 45 & 146 & 170 & 85 & 0.2 & 0.17 & 4.6 x 3.7 & 0.5 x 0.4 & Sep93\\
7742  &  SA(r)b    & +0.7 & +7.5 & 18 &  133 & $-$ &no bar& 0.05 &0& 9.9 $\times$ 9.4 & 1.0 $\times$ 1.0 & Dec03\\
\hline
\end{tabular}}
\end{center}
Notes: Morphological and photometric parameters for the observed sample. Galaxies are listed by NGC number in order of increasing RA (col.~1) with morphological type (col.~2) from de Vaucouleurs et al. (1991; hereafter RC3). Col.~3 lists the offset of the optical center of the \halpha\ image from  the center of the \DensePak\ array; this is not applicable to the TAURUS data for NGC~1300, NGC~4321, and NGC~6951. The nuclear ring inclination $i$ (col.~4) and photometric ring position angle ${\rm \phi_{r}}$ (col.~5) have been derived from \halpha\ imaging by M08). The disk position angle ${\rm \phi_{d}}$ (col.~6) is  from the RC3 with a dash indicating that a ring is circular with no definable position angle. The bar position angle ${\rm \phi_{b}}$ (col.~7) and photometric ring ellipticity ${\rm \epsilon_{r}}$ (col.~8) are from M08. Disk ellipticity ${\rm \epsilon_{d}}$ (col.~9) is from RC3. The radius of the ring, shown as the semi-major axis by the semi-minor axis (col. 10 and col. 11), was derived from the imaging data of M08 using the distance given there. The ring semi-major axis and ring ellipticity values are all consistent with those reported in \cite{C10}, taking into account that the methodology applied was slightly different. Run dates are in col.~12.

\end{table*} 

The kinematic sample was chosen from the imaging parent sample based on unambiguous detection of \hii\ regions delineating the ring, a non-zero inclination ($i$), and a measurable ring position angle (${\rm \phi_{r}}$). Rings too diffuse or small, or with no resolved \hii\ regions detected using the method described in M08 (i.e., weak signal and resolution), were rejected. With the Densepak instrument we cannot detect any \hii\ regions for nuclear rings with angular sizes less than 0$\farcs$5 in radius, however, we have previously unpublished high resolution Fabry-Perot \halpha\ data of NGC~1300 and NGC~6951 from the TAURUS instrument on the 4.2\,m William Herschel Telescope,  which we include in this sample. Similarly, we have used the TAURUS data for NGC~4321 from Knapen et al. (2000), a nuclear ring which meets our criteria but was added to the M08 sample after the DensePak observing runs. Lastly, although not included in the M08 sample, we observed two well 
 known nuclear rings (NGC~2903 and NGC~3351) during the December 2003 run, which we also add to this sample. 

We obtained two-dimensional \halpha\ velocity fields of the sample over two observing runs, in December 2003 and April 2004, using the \DensePak\ fiber-optic array and bench spectrograph on the 3.5\,m WIYN\footnote{Joint facility of the University of Wisconsin-Madison, Indiana University, Yale University, and National Optical Astronomy Observatories} telescope. The \DensePak\ integral field unit (IFU) spectrograph \citep{BW88} consists of 91 fibers bonded into a 7~$\times$~13 staggered rectangular grid that covers an area of $30\times45$\,arcsec of sky with center-to-center fiber (each fiber is 2.82\,arcsec in diameter) spacings of 3.75\,arcsec. Four of the fibers are allocated as sky fibers, and are spaced around the grid roughly 1$\am$ from the center. The spectrograph was configured with the Bench camera, which uses a Tek 2048 CCD (T2KC), with the 860 $l$\,mm$^{-1}$ grating, in the second order, centered at 6575\,\AA. We used a third order blocking filter (GG495). This arr
 angement provided a final spectral coverage of 925\,\AA, from 6085\,\AA\ to 7010\,\AA, at a resolution of 0.45\,\AA\ per pixel. The spectral range included \halpha\ as well as the \nii$\lambda\lambda6548,6583$ and \sii$\lambda\lambda6716,6731$ lines. The effective instrumental resolution was different for the two runs, which is likely due to differences in the instrument setup. For the December 2003 run, the instrumental resolution (FWHM) averaged 0.97\,\AA\ (44\,$\kms$) around \halpha, with the resolution for the April 2004 run 1.31\,\AA\ (60\,$\kms$). 

The array contains six dead fibers which decreases the number of active fibers to 85. To avoid these areas, the projected image of each nuclear ring was placed in the upper half of the fiber array. Because of this method, many of the rings were close to the edge of the field of view (FOV) along the array minor axis, which limits our discussion of the environment exterior to the rings. We used two to four pointings, each of 1800\,s in duration, offset to fill the area of the dead fibers. This method also improved the sampling in some cases. For all exposures, the major axis of the array was aligned along a North-South direction.

The Fabry-Perot observations for NGC~4321 are detailed in \cite{JK00}, with very similar procedures implemented for NGC~1300 and NGC~6951. The pixel size in the reduced Fabry-Perot cubes is 0.27\,arcsec squared, and the separation between velocity channels is 15.7$\kms$. Each cube has 55 planes, thus covering a total velocity range of just over 1000$\kms$. Continuum subtraction was performed using the channels on either extreme end of this range.

\section{Data Reduction}

The raw CCD frames of the DensePak data were zero- and bias-corrected, flat-fielded, cosmic-ray cleaned, and combined using the standard {\sc iraf noao} packages ({\sc zerocombine, ccdproc, flatcombine, cosmicrays}, and {\sc imcombine}, respectively). The combined images were wavelength-calibrated using the {\sc noao iraf} data reduction package {\sc dohydra} to extract the spectrum yielded by each fiber. A CuAr comparison source was used for wavelength calibration. The reduced two-dimensional spectral image contains 89 rows (comprising 85 data fibers and 4 sky fibers), where each row is one pixel in height and corresponds to a single fiber.
 
We created {\sc gipsy}\footnote{Groningen Image Processing System, http://www.astro.rug.nl/gipsy} scripts to identify and then subtract the continuum and sky from the spectrum following the method described in \cite{B05}. Due to pointing uncertainties inherent to the instrument, we established the measured pointing offsets (i.e., the shift from the center of the \DensePak\ fiber array to the ring center) from the observations themselves. We followed the procedure outlined in \cite{SW03}, in which the continuum levels measured from the spectra are compared with levels derived from a reference image with known coordinates.  For our sample we used the $I$-band images from M08. By repeating this process for all positions near the nominal pointing, we created a grid of $\chi^{2}$ values, in which the minimum $\chi^{2}$ corresponds to the measured pointing offset. We estimate that the uncertainty on the derived pointing is approximately 0.5\,arcsec. Because the sky fibers contained
  no \halpha\ emission from the target galaxy, we used the average of the four reserved \DensePak\ sky fibers, each located 1\,arcmin from the center of the array.

To combine the pointings into a sparse velocity field, we first fitted Gaussians to the continuum- and sky-subtracted \halpha\ emission line profiles, using a 3$\sigma$ threshold to avoid spurious detections. The fitted velocities derived for each fiber were placed in a map at corresponding fiber positions. The observed radial velocities were corrected to heliocentric velocities. In the cases of multiple velocity measurements for the same pixel in the velocity field, we averaged the measured velocities. To aid in the visual representation of the sparse velocity field, we also constructed interpolated smoothed velocity field maps. We replaced each data point in the sparse velocity field with a Gaussian beam of 5.5\,arcsec and weighted the overlapping sections between the points by the relative intensities of the overlapping Gaussians. The resulting smoothed velocity fields are shown in Figure~\ref{fig:vf} (left, panel~1). Because of interpolation effects these contiguous velocity fields may be uncertain, especially near the edges. Any quantitative interpretation comes from the sparse velocity maps, which are not interpolated.  
 
\begin{figure*}
\setcounter{figure}{0}
\includegraphics[angle=-90, width=6.5in]{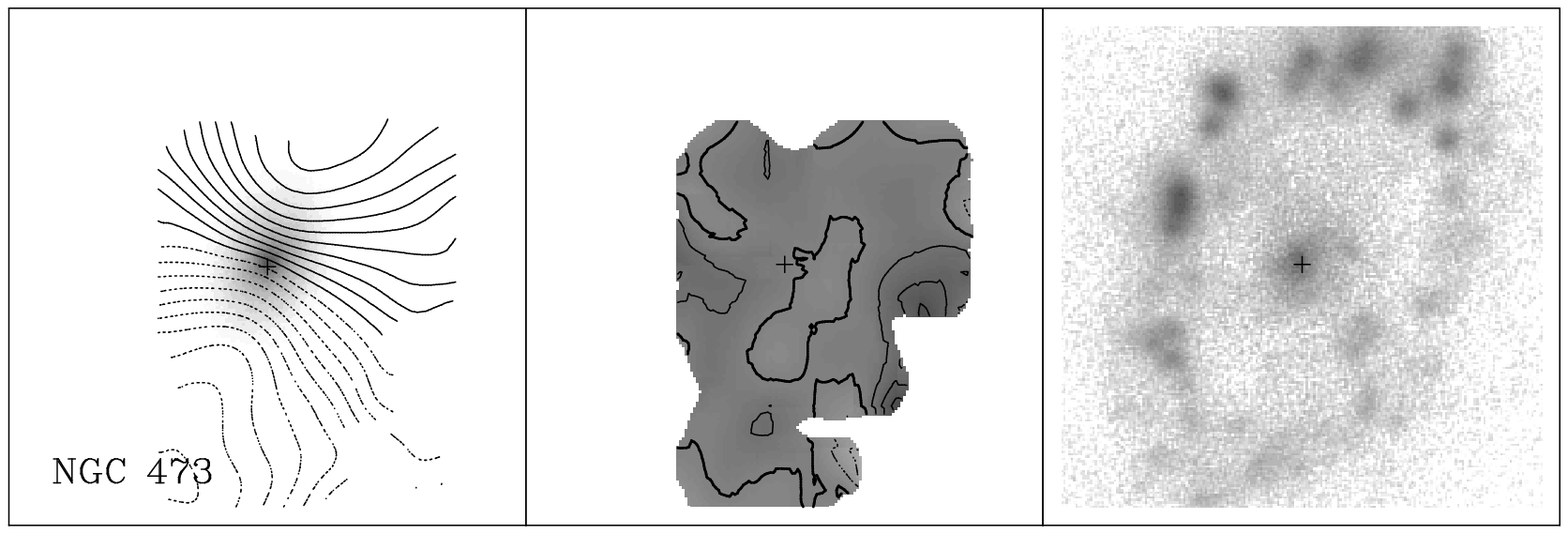}
\includegraphics[angle=-90, width=6.5in]{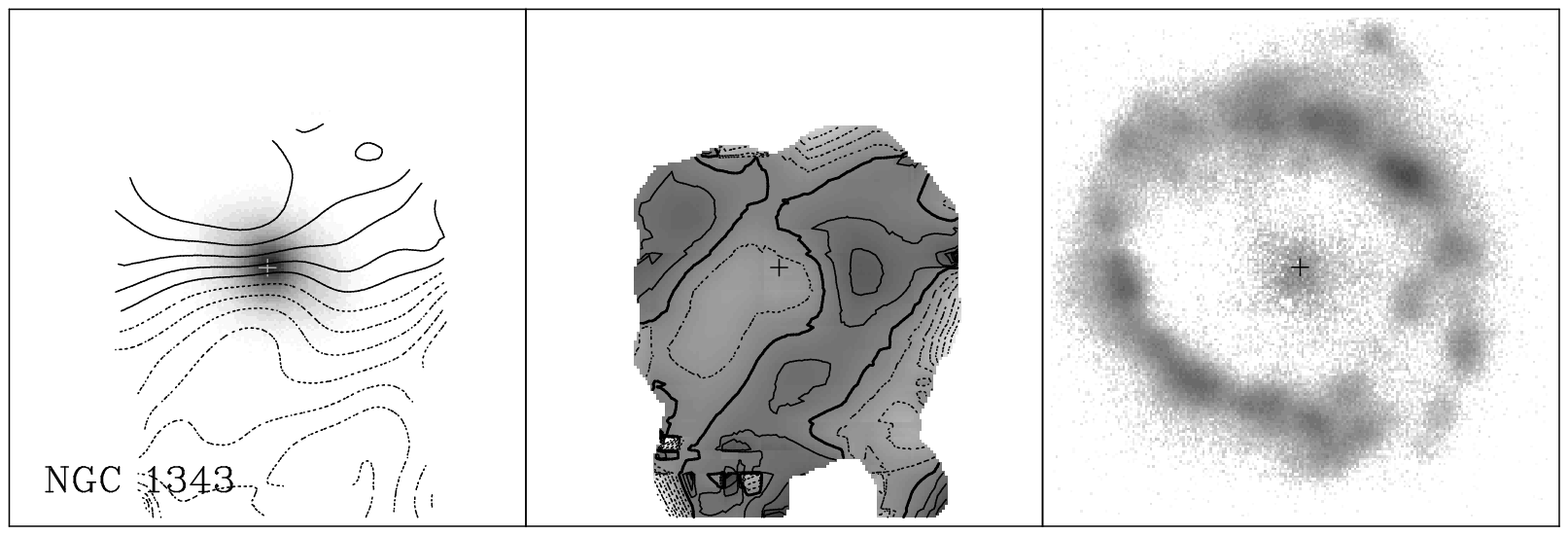}
\includegraphics[angle=-90, width=6.5in]{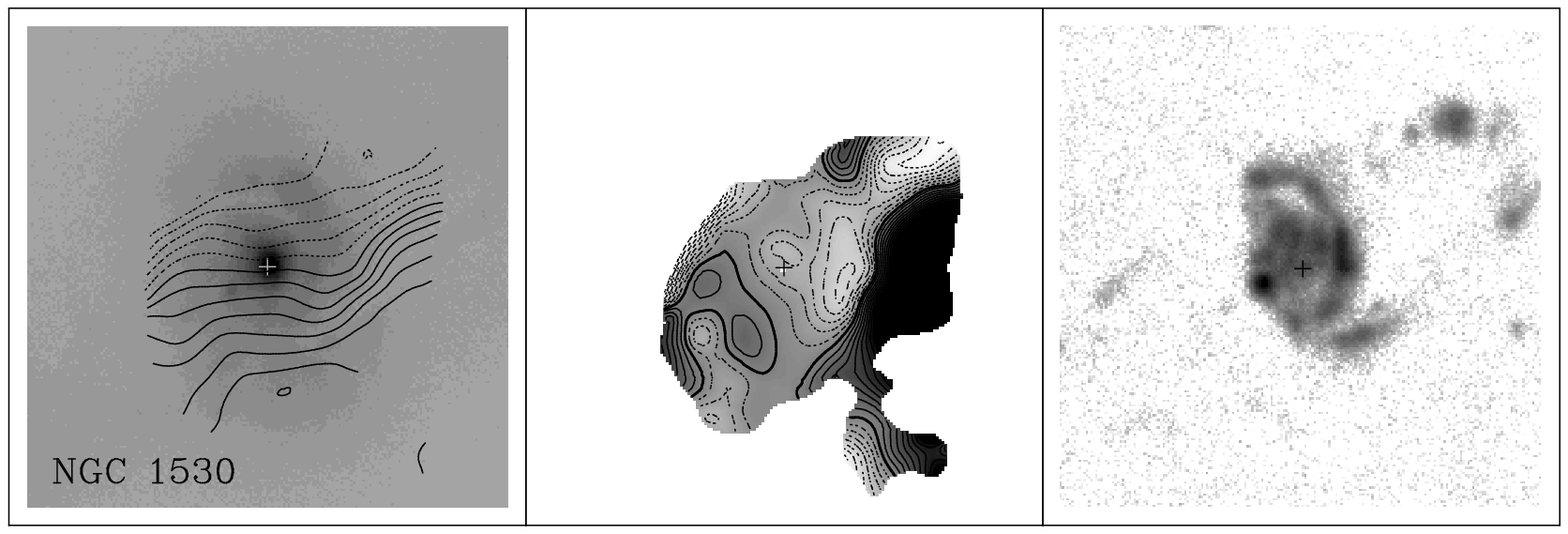}
\caption[]{Results from the DensePak sample. From left to right, $I$ band image from M08 with velocity field overlaid in contours, residual velocity field with contours, and \halpha\ emission image from M08. The center of the galaxy is indicated with a cross in each panel. Contour levels in the left panel are spaced by 5 $\kms$ with dotted contours representing velocities less than the systemic velocity and solid contours representing those velocities greater than the systemic velocity. Residual velocity contours and grayscales within the rings generally range from $-$10 to 10 $\kms$ except for NGC~1530 whose contours range from $-$20 to 20\,km\,s$^{-1}$, and NGC~473 and NGC~7742 which both have residual ranges from $-$5 to 5\,km\,s$^{-1}$. N is up, E to the left. The size of the field shown is 10\,arcesc (RA) by 40\,arcsec (DEC) for all galaxies.}
\label{fig:vf}
\end{figure*} 
 
\begin{figure*}
\setcounter{figure}{0}
\includegraphics[angle=-90, width=6.5in]{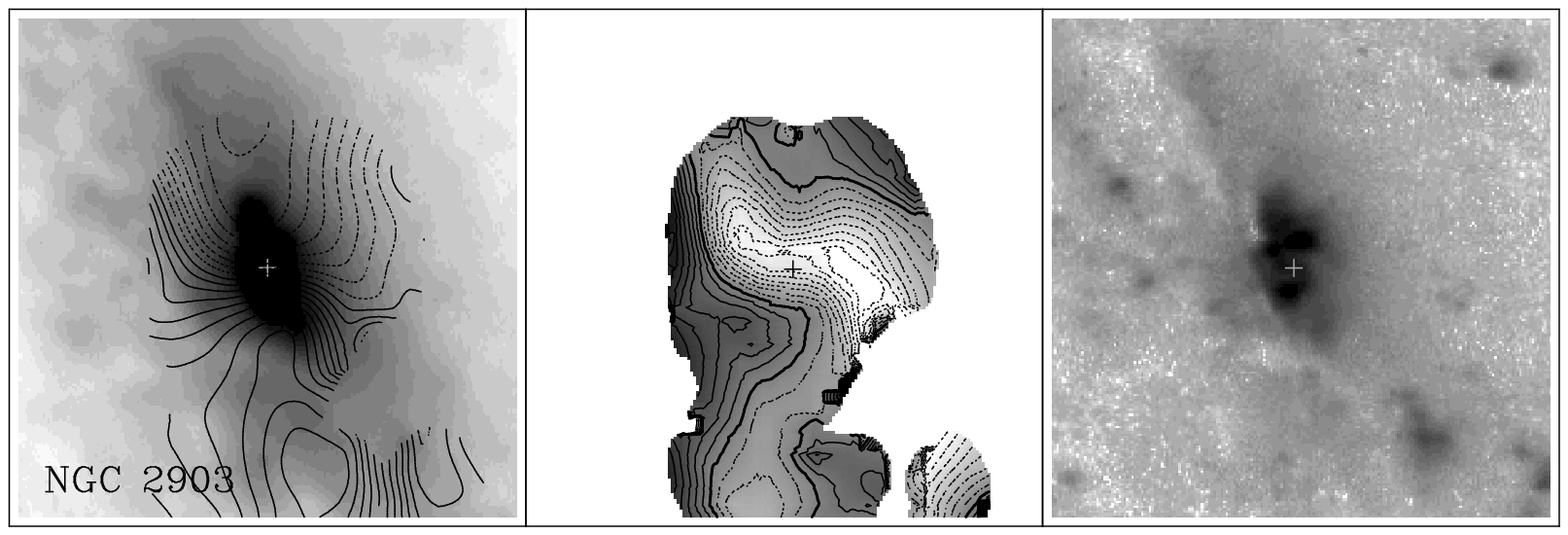}
\includegraphics[angle=-90, width=6.5in]{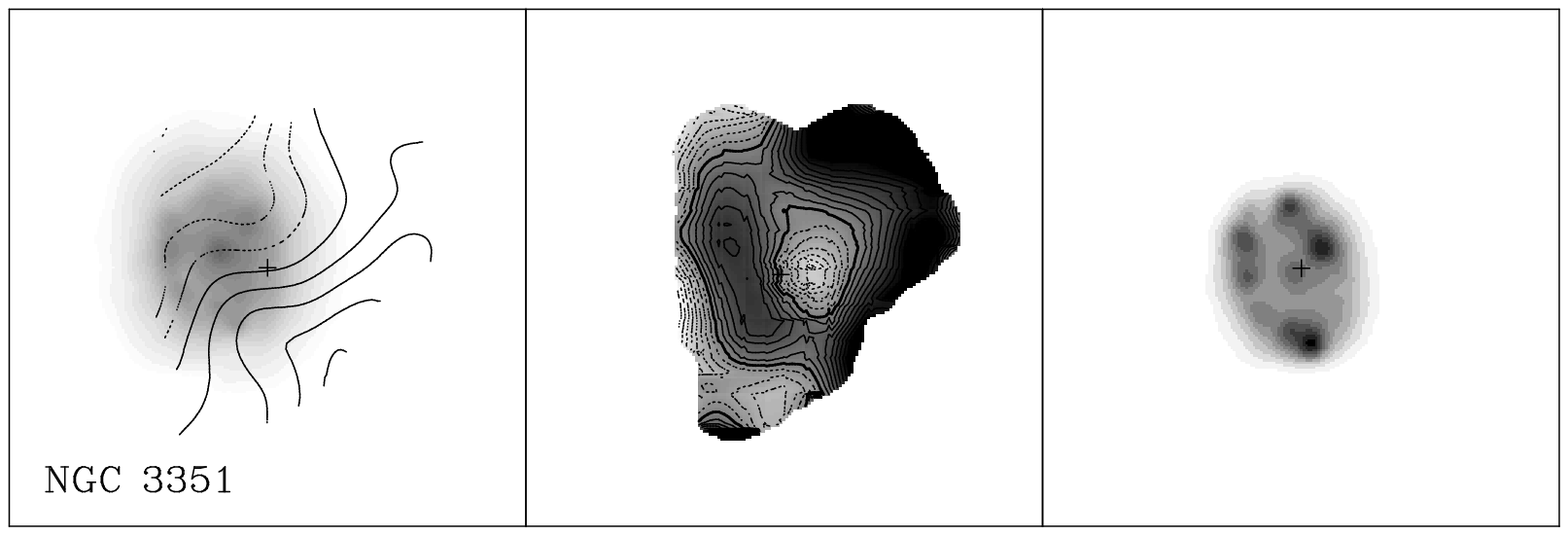}
\includegraphics[angle=-90, width=6.5in]{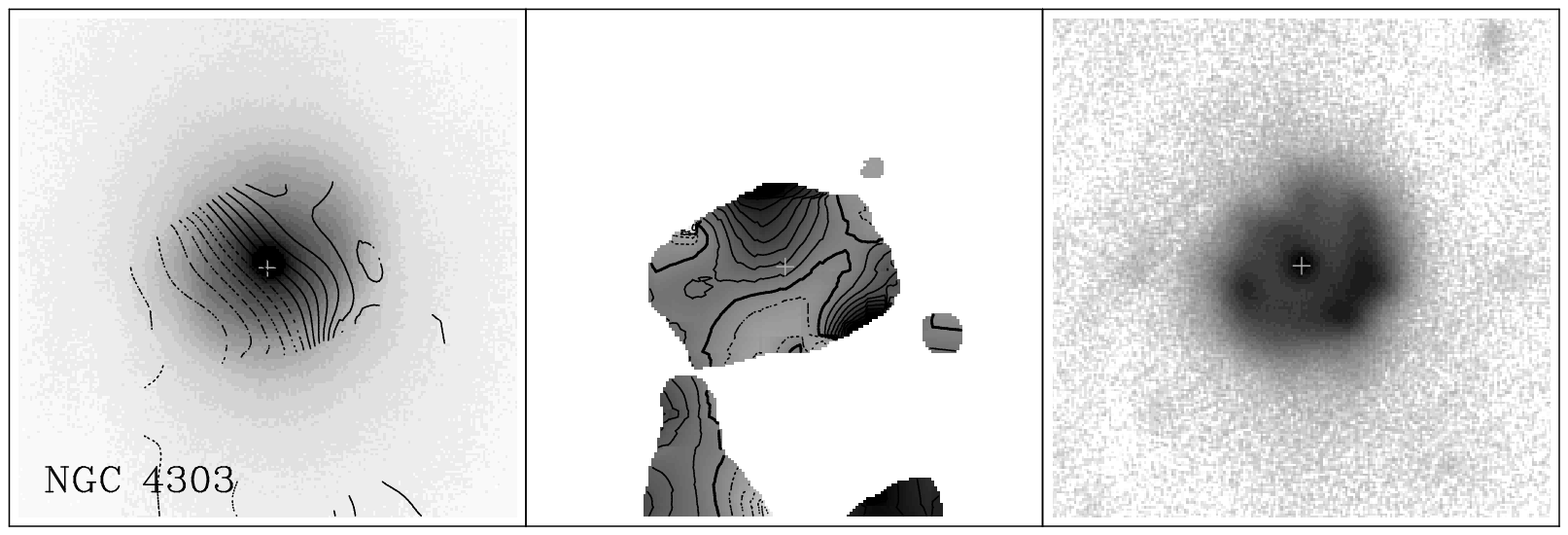}
\includegraphics[angle=-90, width=6.5in]{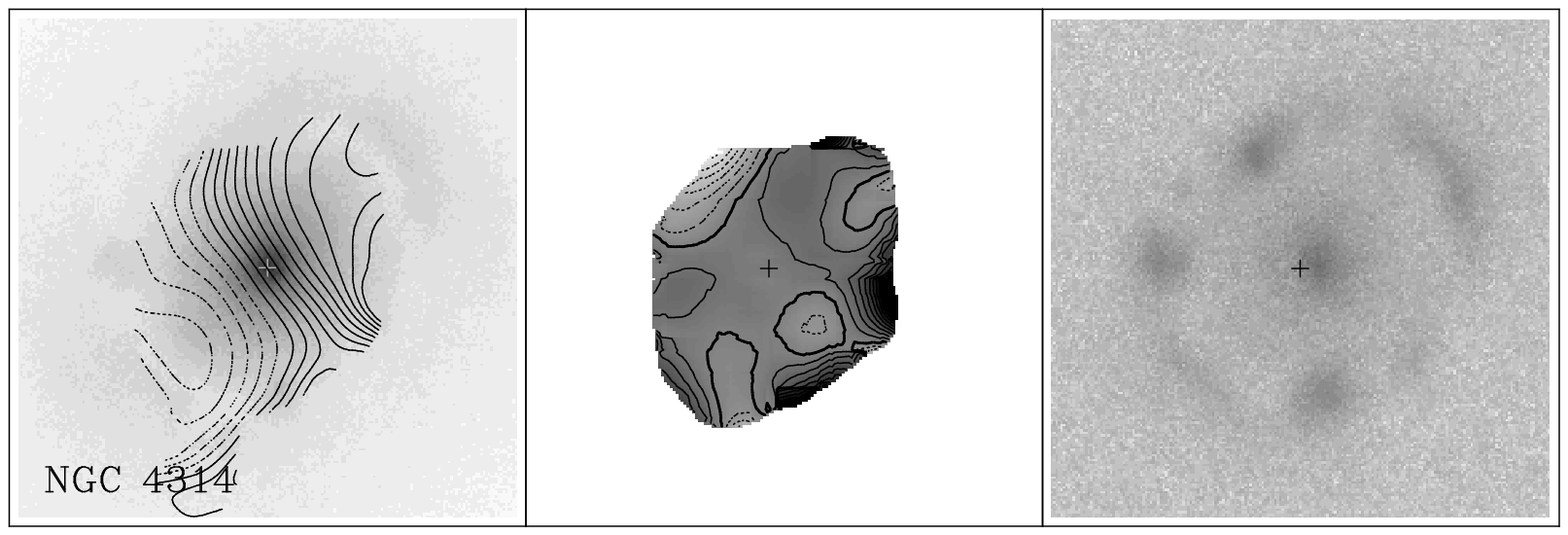}
\caption[]{continued.}
\end{figure*} 

\begin{figure*}
\setcounter{figure}{0}
\includegraphics[angle=-90, width=6.5in]{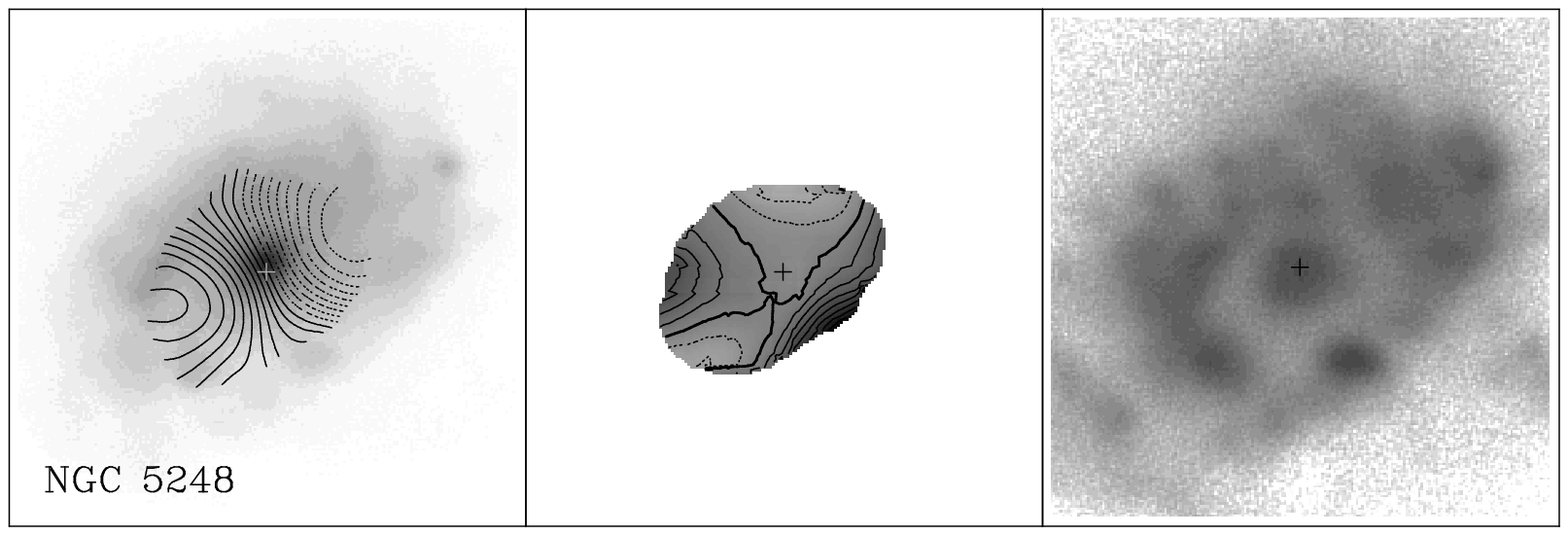}
\includegraphics[angle=-90, width=6.5in]{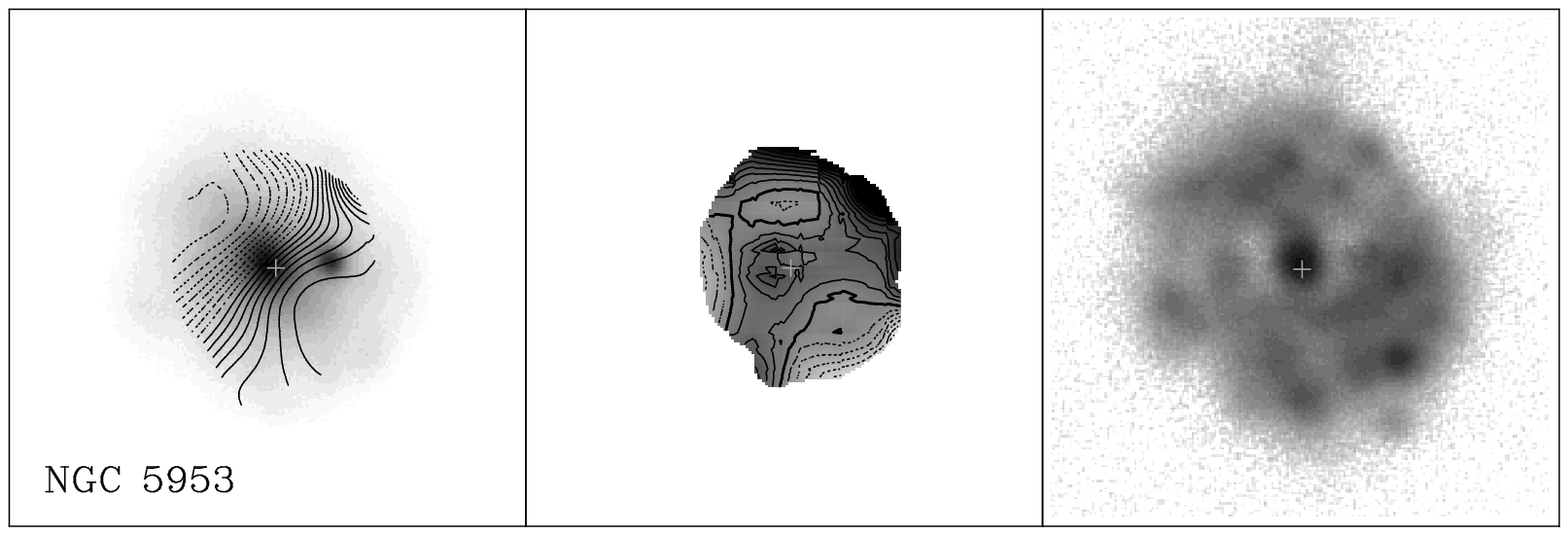}
\includegraphics[angle=-90, width=6.5in]{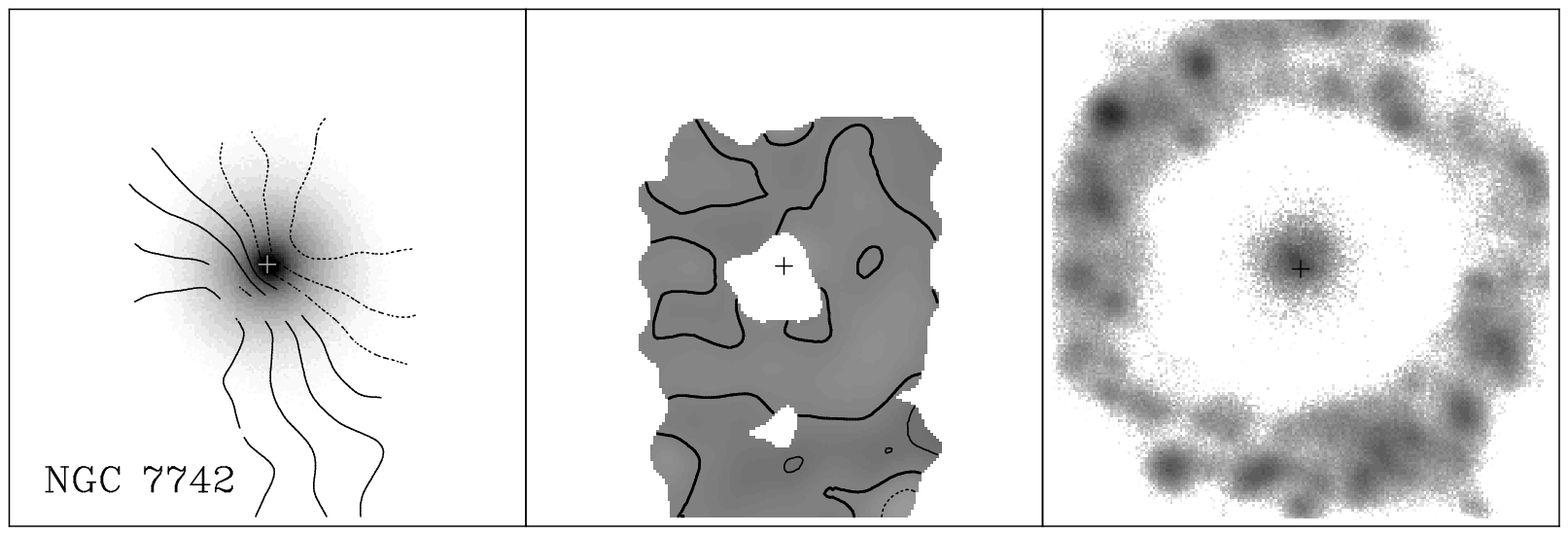}
\caption[]{continued.}
\end{figure*} 
 
\begin{figure*}
\setcounter{figure}{1}
\vspace{-1cm}
\includegraphics[trim = 1in 4.1in 1in 4.6in, clip,scale=1.5, width=6.5in]{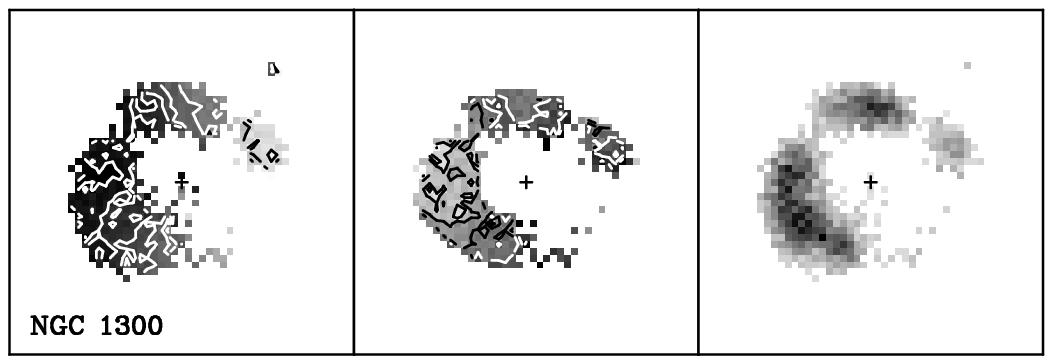}
\vspace{-1cm}
\includegraphics[trim = 1in 4.1in 1in 4.6in, clip,width=6.5in]{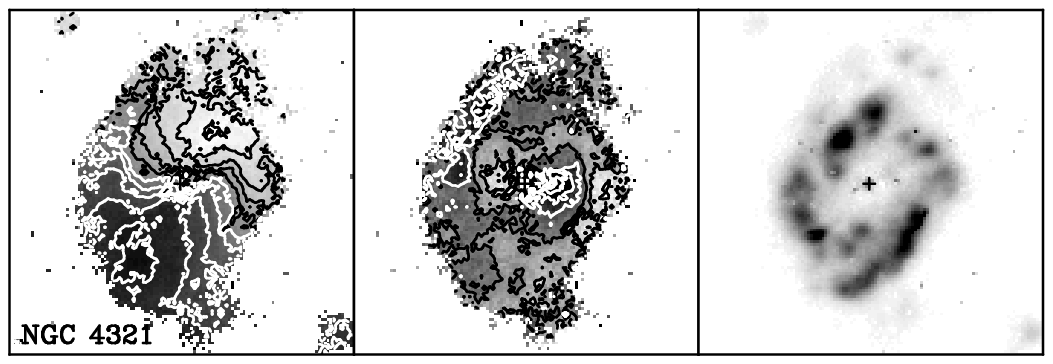}
\includegraphics[trim = 1in 4.1in 1in 4.6in, clip,width=6.5in]{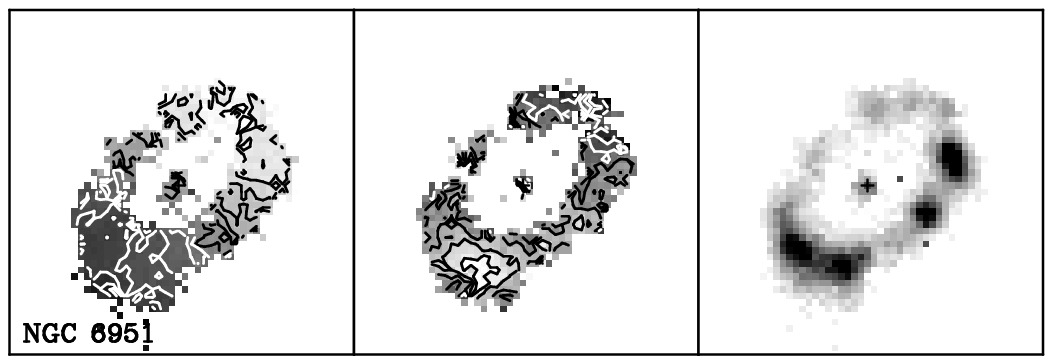}
\caption{As Fig.1, now for the three galaxies observed with the TAURUS Fabry-Perot instrument. Panels are, from left to right, velocity field (grayscale and contours), result after subtracting a model velocity field based on the rotation curve from the observed velocity field, and \halpha\ integrated intensity. The center of the galaxy is indicated with a cross. The field of view shown is 15, 35, and 15\,arcsec square for NGC~1300, NGC~4321, and NGC~6951, respectively. Contour levels for the first (leftmost) panel are for NGC~1300 from 1472 to 1552\,km\,s$^{-1}$ (white contours) and from 1572 to 1672\,km\,s$^{-1}$ (black); for NGC~4321 from 1480 to 1560 (white) and from 1580 to 1680\,km\,s$^{-1}$; and for NGC~6951 from 1300 to 1440 (white) and from 1460 to 1600\,km\,s$^{-1}$---all contours are spaced by 20\,km\,s$^{-1}$ and the grayscale range is the same as that spanned by the contours. In panel 2, contour levels for all galaxies are from $-$45 to 0\,km\,s$^{-1}$ (white) and from 15 to 45\,km\,s$^{-1}$ in steps of 15\,km\,s$^{-1}$; the grayscale range is from $-$50 to 50\,km\,s$^{-1}$ for NGC~1300 and NGC~6951 but from $-$30 to 30\,km\,s$^{-1}$ for NGC~4321. For all three galaxies, the grayscale in panel 3 indicates the range of values in instrumental units.}
 \label{fig:vf-FP}
 \end{figure*} 
 
The Fabry-Perot data were wavelength and phase calibrated, with sky and continuum emission subtraction performed, as described in detail in \cite{JK00}. The resulting reduced data cubes were used to produce \halpha\ velocity field maps using {\sc gipsy}, and following the prescriptions given by \cite{JK97} and \cite{JK00}. The resulting velocity fields are shown in Figure~\ref{fig:vf-FP}.

 
\section{Analysis}

\subsection{Rotation Curves}

Rotation curves are powerful tools for revealing intrinsic kinematic characteristics of galaxies, including those of the nuclear rings they may host. In our sample, the range in radius spanned by the rotation curves is small when compared to the overall disk size, but this is a result of the small field of view of the instrument and of the sensitivity limit. 

\begin{figure*}
\begin{center}$
\begin{array}{c}
\includegraphics[width=6in]{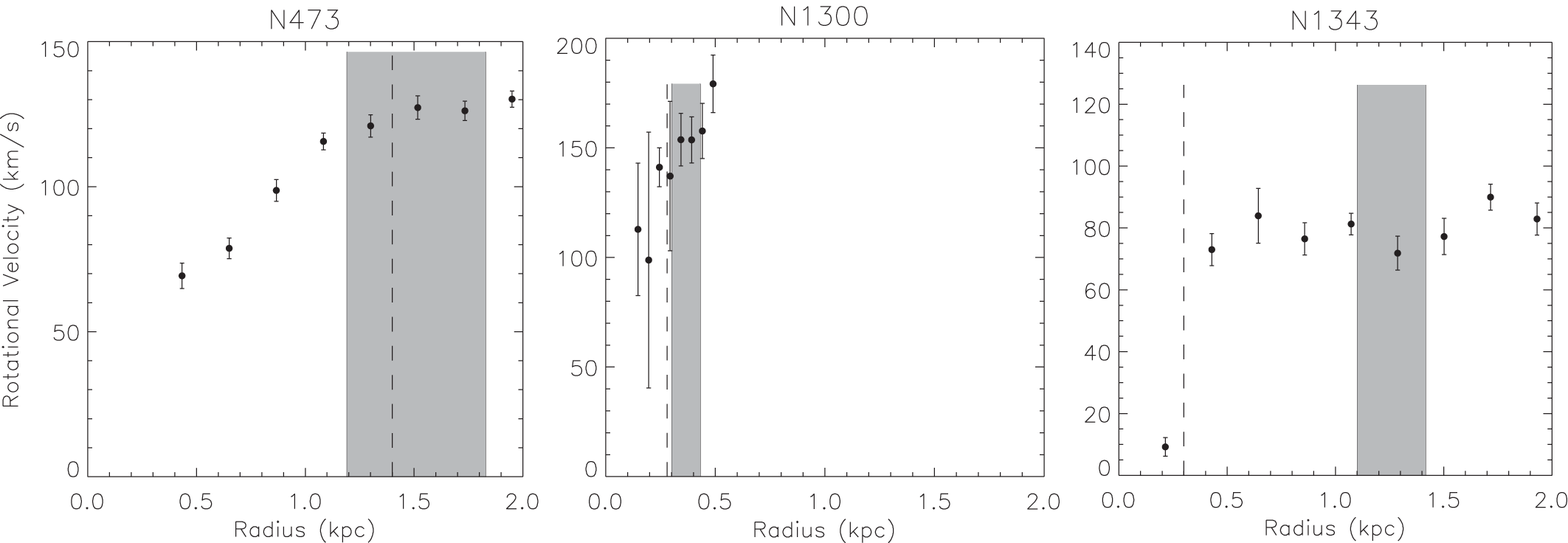} 
\\ \includegraphics[width=6in]{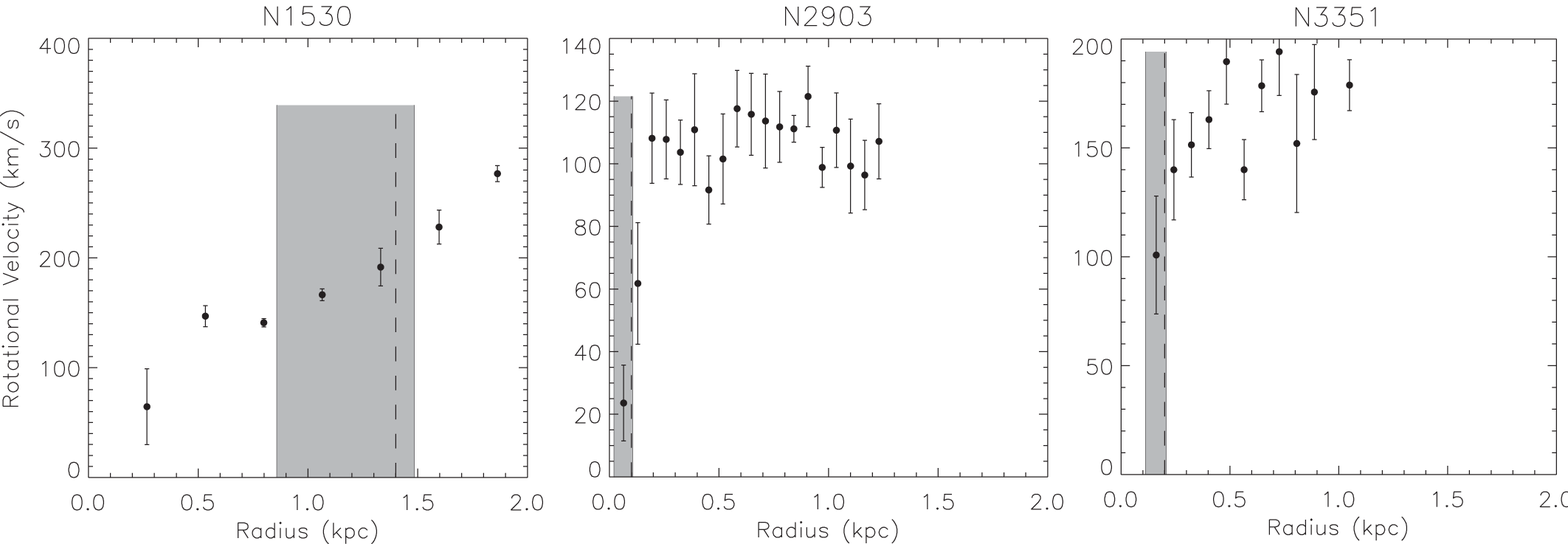} 
\\ \includegraphics[width=6in]{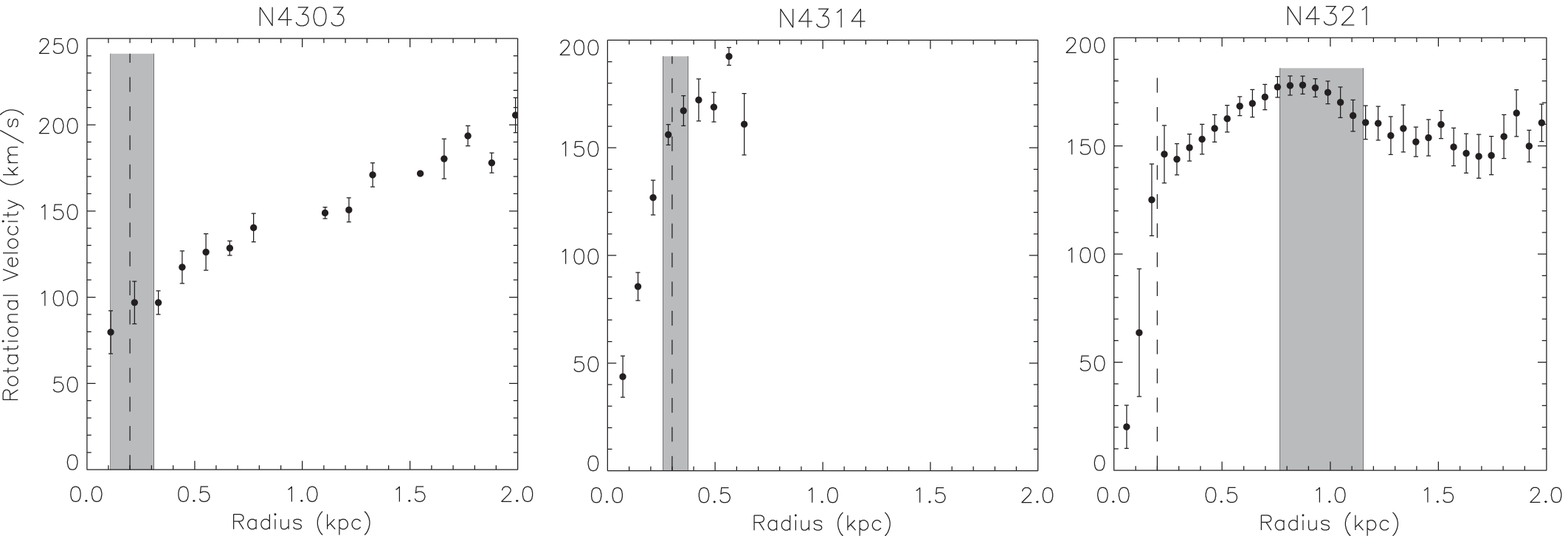} 
\\ \includegraphics[width=6in]{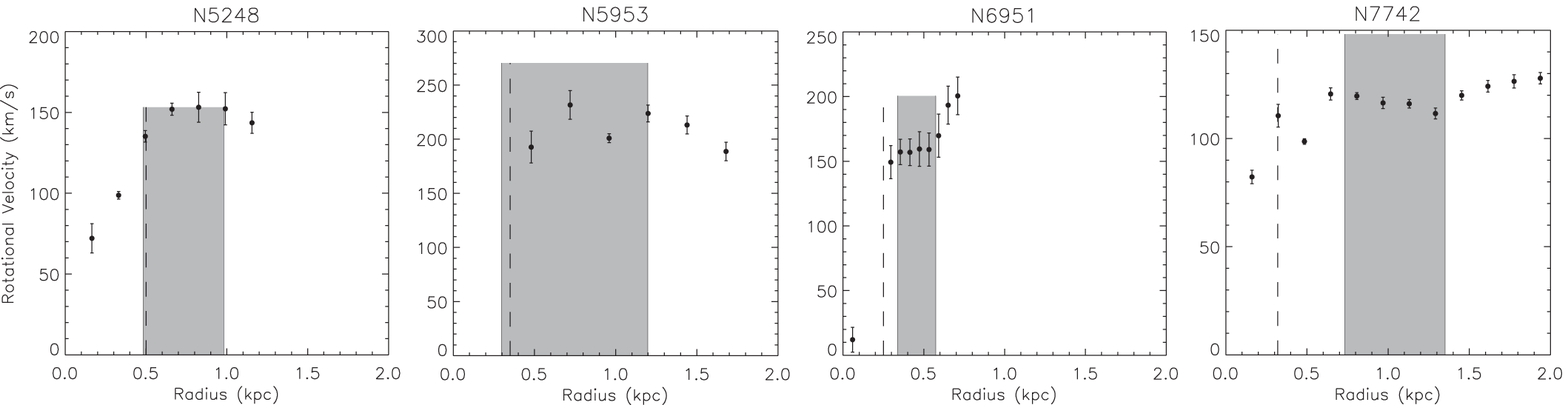} 
\end{array}$
\end{center}
\caption{Rotation curves with the radial range (i.e., width) of the nuclear rings shadowed. The turnover radius is denoted by the dashed vertical line. See Table~\ref{tab:rc-parms} for ring widths.}
\label{fig:rc}
\end{figure*}

For each galaxy we derived the rotation curve (Figure~\ref{fig:rc}) by fitting tilted rings to the sparse velocity field through an iterative process (see \cite{B89} for procedural details).  We chose the incremental radius for each concentric ring to be 1.5\,arcsec, which is approximately half of the width of a given fiber. We supplied the initial input values for the nuclear ring center, inclination, and position angle, as measured from $I$-band images (M08). We derived the initial center position by taking the point where the $I$-band emission is the highest near the galaxy center. The input value for $v_{sys}$ in the fit to the velocity field is from the NASA/IPAC extragalactic database (NED). One exception is NGC~5953, whose systemic velocity was derived from the data due to a 26$\kms$ difference with the NED value. For the initial run, although we supplied input values, all parameters were left free. We first determined the center position and then fixed only that value for the second iteration to then determine the systemic velocity. Once the center and systemic velocity were fixed we determined the inclination and position angle (third and fourth iterations, respectively). With the four values then fixed, we fit the rotational velocity. We gave more weight to positions along the major axis, since those data points carry most of the information about rotational velocities. In most cases we used the inclination of the outer disk for the final fit. However, in the cases of low inclination (below 25\deg), we adopted the photometric orientation parameters from M08 as initial inputs to the routine since the fitting routine produced inclinations with high uncertainties. Even with such low inclinations the position angle was determined accurately with a tilted ring fit. Because our data sets span a small range in radius, it was not feasible to make sophisticated fits to the velocity fields. 

The uncertainties on the velocities derived from the line fits range from 0.5$\kms$ to 5$\kms$. An additional source of uncertainty is due to the unknown location of the Halpha emission within the area seen by each fiber. The amplitude of this uncertainty depends on the velocity gradient across the face of the fiber. We estimated this gradient for all fibers using the derived rotation curves, and found that a value of 8$\kms$ is representative of a 1$\sigma$ uncertainty.

Rotation curves for the Fabry-Perot data were similarly generated, but here, the smaller pixel size and full spatial sampling allowed us to fit the rotation curves with radial steps of 0.5\,arcsec from the center outward.

\subsection{Residual Velocity Fields} 

We constructed residual velocity maps (Figure~\ref{fig:vf}, panel~2) for the velocity fields (both the sparse and interpolated versions for the DensePak data) by subtracting a smooth model velocity field, constructed from the rotation curve and the assumed orientation parameters. As with the interpolated velocity fields, the interpolated residual maps should only be used for general impressions. These maps, however, are useful because they give a better sense of patterns in the image.

\subsubsection{Within the rings}
We find that all of the nuclear rings in our sample have low residual velocities (both approaching and receding) near the central ridge of the ring (ranging from $\sim$5$\kms$ to $\sim$20$\kms$). The nuclear rings in NGC~473 and NGC~7742 are not resolved enough to comment on here, as the associated residuals do not overcome the uncertainties associated with the measured velocity, wavelength calibration, and position of the \halpha\ peak within a given fiber beam. Patches of approaching and receding residual velocity peaks of $\sim$10$\kms$ occur in the nuclear rings of NGC~1343, NGC~4314, and NGC~5953, and reach to near 20$\kms$ in the rings of NGC~1530 and NGC~5248. Since these values are close to the combined measurement uncertainty, they can provide no firm evidence for the existence and origins of non-circular motions (in NGC~4321 the situation is better, as discussed in detail by Knapen et al. 2000).

A comparison of the kinematic and photometric centers of the sample reveals no strong asymmetries. The degree to which the centers differ is proportional to the lopsidedness of the potential, and thus a good indicator of the amount of non-circular motion \citep{F94}.  After computing the difference between the kinematic and photometric center positions (see Table~\ref{tab:kin}), we see that the center positions agree to within the uncertainties of the photometric center measurements. This agreement, in combination with the small noncircular motions, suggests the nuclear rings are circular in nature. 

\begin{table*} 
\begin{center}
\caption[]{Kinematic parameters for the observed sample}
\label{tab:kin}
\begin{tabular}{lcccccccc}
\hline \hline
NGC & $V_{\rm sys}$ & \multicolumn{2}{c}{Offset} & ${\rm \phi_{kin}}$ & ${\rm \mid \phi_{kin} - \phi_{phot}\mid}$ & ${\rm \mid \phi_{d} - \phi_{kin}\mid}$ & \multicolumn{2}{c}{${\rm \mid Ctr_{kin} - Ctr_{phot}\mid}$} \\
& & $x$ & $y$ & & & & $x$ & $y$\\
 & (km\,s$^{-1}$) &($\as$) & ($\as$) & ($\deg$) & ($\deg$) & ($\deg$)  & ($\as$)&($\as$)\\
(1) & (2) & \multicolumn{2}{c}{(3)} & (4) & (5) & (6) & \multicolumn{2}{c}{(7)} \\
\hline
~473  &   2135  & $-$5.0 & +7.5   & 156 &  2 & 3 & 0.6 &  0.3\\
1300  &     505  &  n/a  &  n/a   & 80   &  55 & 26 & n/a &n/a \\  
1343  &   2215  & $-$5.3 & +6.5   & 7    & 53 & 73 &  0.9  & 1.3\\
1530  &   2450  & $-$1.6 & +11.0 & 5   & 20 & 3 &  0  & 0.3\\
2903  &     560  & $-$4.0 & +7.4   & 5    & 3 & 12 &  2.8 & 0.9\\
3351  &     764  & $-$4.2 & +7.0   & 30 & 10 & 17 &  0 & 0.9 \\
4303  &   1560  & $-$3.4 & + 7.5   & 117 & 29 & $-$ &  0.1 & 0\\
4314  &     979  & $-$4.0 & +5.5   & 122  & 13 & $-$ &  1.0  & 0.8\\
4321 &    1578  &  n/a    &   n/a  & 150   &  20 & 120 &n/a  &n/a \\  
5248  &   1149  & $-$2.9 & +6.3   & 115  & 0  & 5 &  0.4  & 0.2\\
5953  &   1991  & $-$6.0 & +6.7   & 49    & 57 & 120 &  0  & 0.1\\
6951  &   1423  & n/a    &  n/a    & 136   &  10 & 106 & n/a &n/a \\  
7742  &   1658  & +0.7 & +6.6  & 131 & 2  & $-$ & 0  & 0.9\\
\hline
\end{tabular}
\end{center}
Notes: NGC numbers of the target galaxies are listed in order of increasing RA (col.~1). Systemic velocity (col.~2), nuclear ring kinematic center position with respect to the center of the \DensePak\ array (col.~3) and kinematic position angle (col.~4) were derived using the {\sc gipsy} {\sc rotcur} task. Typical uncertainties for $V_{\rm sys}$ are $\pm8\kms$, center position $\pm0.09$\,arcsec, and position angle $\pm5\deg$. The difference (absolute value) between the kinematic and photometric position angles (see Table~\ref{tab:phot}) is in col.~5, with the difference between kinematic and disk position angles in col.~6. The difference between kinematic and photometric centers is in col.~7.
\end{table*}

\subsubsection{In the vicinity of the rings}

For some of the galaxies in our sample, our observations cover part of the velocity field outside of the nuclear ring as well. This allows us
to study the interaction between the bar and the nuclear ring in those cases.

Sufficient, albeit limited, \halpha\ radial coverage allows us to comment on the velocities of the barred galaxies near the outer edge of the nuclear ring, perpendicular and parallel to the bar major axis. In all cases, we can see the influence of the strong bars as they create high residual velocities. However, once they interact with the nuclear ring outer boundary, residual velocities sharply decrease and plateau within the nuclear rings. For NGC~4314, we see ordered velocity patterns increasing near the outer edge of the ring (both approaching and receding) close to both sides of the bar minor axis, where the bar dust lanes merge with the ring. Residuals approach 60$\kms$ in both directions, with a strong velocity transition zone near the outer boundary of the nuclear ring, where values reach 90$\kms$ (approaching) at the edge of the FOV. Although the spatial sampling is poor at the image edge, our observations do agree with velocity values from \cite{B96}. 

The residuals for NGC~5248 near the western side of the bar minor axis reach 40$\kms$, with a weaker-defined, although evident, transition zone as radii approach the outer ring boundary. These results are consistent with the two-dimensional CO velocity field  presented by \cite{J02}. In both galaxies, good agreement with the literature adds confidence that the results we see at the field edges are not artificial, even though our data are spatially sparse. 

\begin{figure}
\includegraphics[height=3in,width=3.1in]{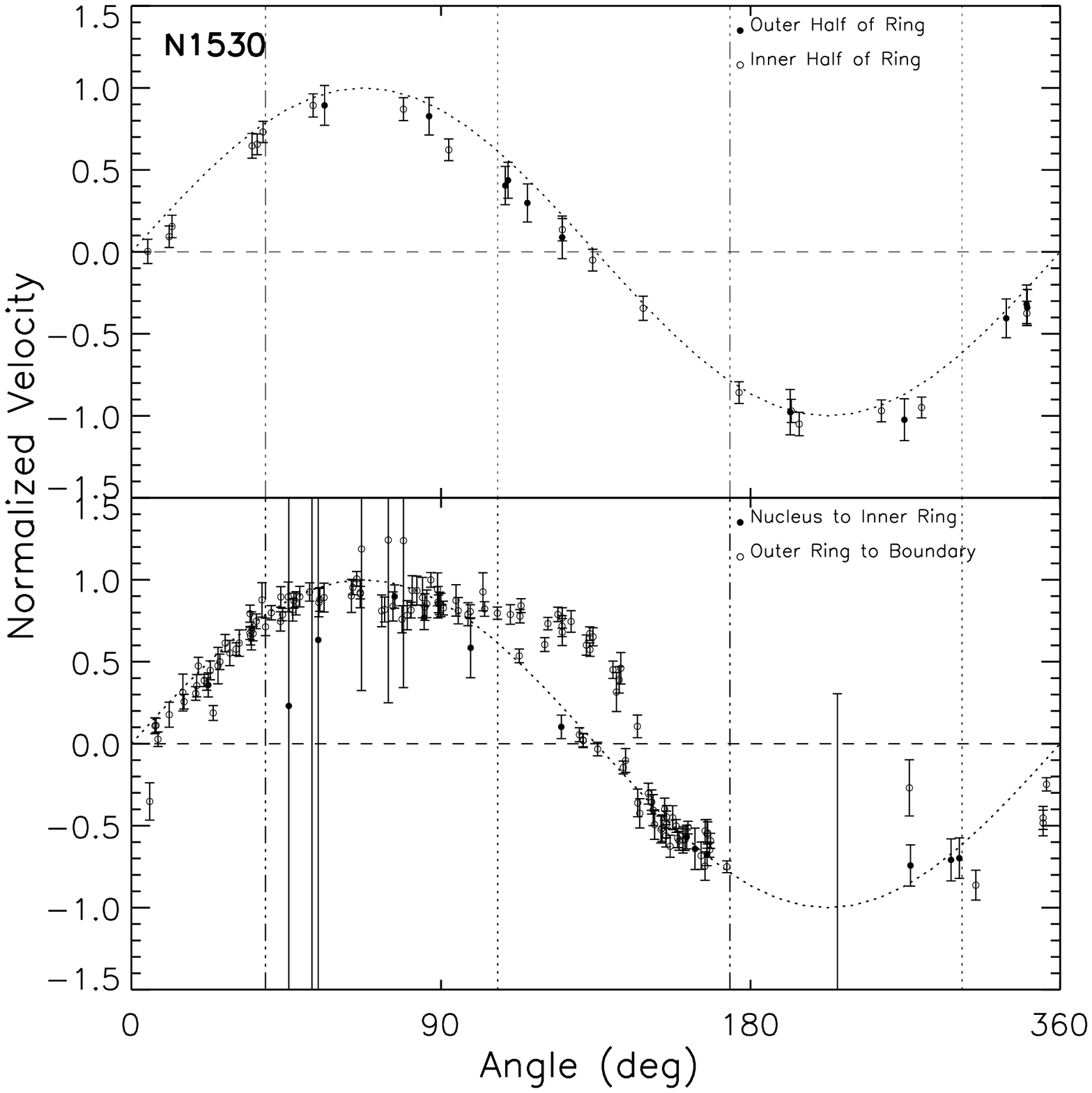}
\caption{Velocity curve for NGC~1530 at radii within (top plot) and exterior to (bottom plot) the nuclear ring. The radial coverage within the nuclear ring is further divided to show velocities along the inner and outer halves of the ring. Outside of the nuclear ring, velocities from the nucleus to the ring inner edge are plotted, as well as those from the ring outer edge to the radial limits of the FOV. The velocity is plotted with respect to the systemic velocity, normalized by the rotation velocity, and adjusted by removing the inclination in order to allow comparison with the curve representing circular motion. The bar major axis (solid lines) and minor axis (dotted lines) are drawn for context. Note that there is a phase shift of 90$\deg$, plus the difference between the kinematic position angle of the velocity data points and the photometric position angle used to create the model curve, to allow the cross-over point to occur at 180$\deg$. The area azimuthally located b
 etween the bar major and minor axes shows the largest velocities of the sample, with deviations reaching 110$\kms$. } 
\label{fig:n1530}
\end{figure}

The residual map of NGC~1530 provides us with the clearest picture of the interaction between the bar and the ring. In Fig.~\ref{fig:n1530} we see large velocity excesses ($\sim$110$\kms$) to the west of the nuclear ring, near its exterior edge. The high residuals are associated with the large scale bar. The strong residuals sharply decrease to 50$\kms$ over a small transition zone near the outer boundary of the nuclear ring. Although we see no obvious residual patterns near the bar major or minor axes of the nuclear ring, we note that the largest residual velocities ($\sim$120$\kms$) across our sample occur in this galaxy, between the bar major and minor axes, on the west side just exterior to the ring outer boundary. A closer look in the vicinity radially shows a very sharp transition at the location of the nuclear ring, where residual velocities drop to 10$\kms$. Zurita \ea\ (2004) observe a nearly identical pattern in the corresponding \halpha\ map of the circumnuclear environment. 

The TAURUS data of NGC~4321 used here were analyzed and discussed in detail by \cite{JK00}. In the residual velocity map, they highlighted the contributions from the inner part of the bar and the spiral arm fragment coming into the nuclear (pseudo-)ring from the main part of the bar. These result in a broadly symmetric pattern of positive and negative residual velocities, mapping out the non-circular motions due to bar and spiral streaming motions.  

\section{Results}

\subsection{Circular Nuclear Rings}

\begin{figure}
\center
\includegraphics[width=2.4in]{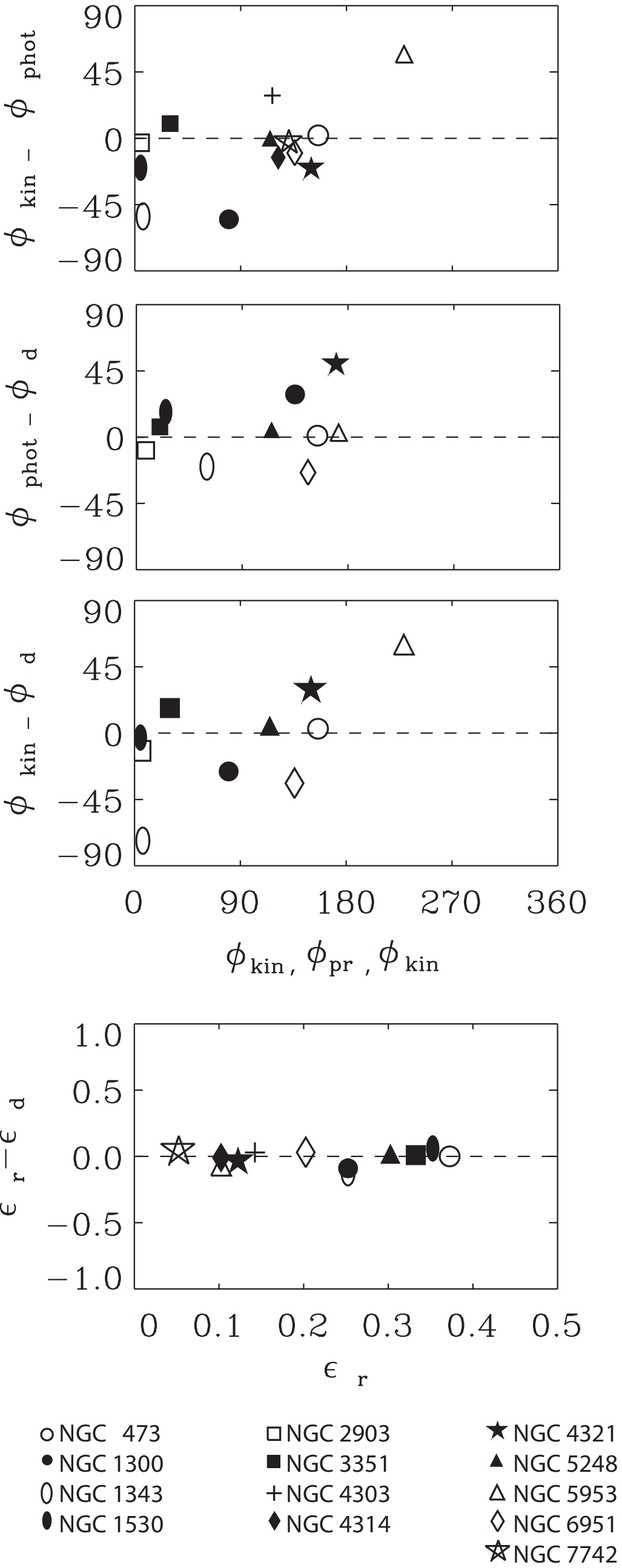}
\caption{Comparisons for ring kinematic vs. ring photometric (top), ring photometric vs. disk photometric (middle), and ring kinematic vs. disk photometric (bottom) position angles. Ellipticity comparison of the ring and disk is also plotted. The disk position angle is not given in the RC3 for NGC~4303, NGC~4314 and NGC~7742. The mean error on the disk position angle is unknown, but the error increases as the ratio of isophotal major to minor axes decreases ($\log R_{25}$). We assigned a 5$\deg$ error for the smallest ratio in the sample and proportionally computed the other errors. Errors are smaller than the size of the graphic icons of each galaxy.} 
\label{fig:pa}
\end{figure}

For most of the sample, we find that the nuclear rings are nearly circular and in the same plane as the disk. The shape and orientation with respect to the host disk were deduced by comparing the position angle (of the nuclear ring, $\phi_r$ and disk, $\phi_d$) combined with the ellipticity of each nuclear ring, $\epsilon_r$ to that of its host disk, $\epsilon_d$. We used the photometric position angles and ellipticities of the nuclear rings from M08. The  kinematic position angles of the nuclear rings were extracted from the rotation curve generation. In Figure~\ref{fig:pa}, we compare the various position angles and see that in most cases the photometric and kinematic position angles are similar. The same can be said for both the photometric and kinematic position angles with respect to the galactic disk position angle. This finding, coupled with the result from M08 of a similar relationship with respect to the disk and nuclear ring ellipticities (fourth panel), corroborate
 s our idea that nuclear rings are circular and oriented as their host disk.

We note three cases that are significantly above the scatter and thus do not meet our criteria of a round in-plane nuclear ring:  NGC~1300, NGC~1343, and NGC~5953. The ring in NGC~1300 is not fully sampled azimuthally (see Figure~\ref{fig:vf-FP}), which adds a high level of uncertainty for the photometric position angle and ellipticity. NGC~5953 is nearly face on and morphologically patchy and round along the line of sight, a combination that creates high position angle uncertainty. 

The residual maps (Fig.~\ref{fig:vf} and Fig.~\ref{fig:vf-FP}, panel~2) also indicate that within the uncertainties of our observations (see Section~4.1), the nuclear rings are circular. There may be noncircularites at a lower level which may be related to the local underlying potential (as noted for NGC~4321 - \cite{JK95}; NGC~1068 - \cite{SC00}; and NGC~5383 - \cite{D77, Sh00}). 

\subsection{Mass Concentration and Nuclear Ring Size}

\begin{table*} 
\begin{center}
\caption[]{Nuclear ring and bar properties.}
\label{tab:rc-parms}
\begin{tabular}{lcccccccc}
\hline \hline
NGC & {\rm Nuclear} & ${\rm Rel Ring}$ & ${\rm Ring}$& ${\rm Q_{\rm g}}$ & Compactness & $R_{\rm t}$ &  Disk\\
& Activity & Size & Width & & &  & sma\\
& & & (kpc) & & ($10^4$\,(km\,s$^{-1}$)$^2$\,kpc$^{-1}$) & (kpc) & (kpc)\\
(1) & (2) & (3) & (4) & (5) & (6) & (7) & (8) \\
\hline
~473   & $-$ & 0.20 & 0.62 &0.12& 1.1 & 1.4  & 7.4\\
1300   & $-$ & 0.02 &0.13& 0.54& 16.5 & 0.31 & 18.2\\
1343   & $-$ & 0.11 &0.31 & 0.15 & 2.3 & 0.3 &11.1 \\
1530   & $-$ &  0.05 & 0.62 & 0.73 & 8.2 & 1.4 &24.5\\
2903   & H\,{\sc ii} & 0.08 & 0.01 & 0.27 & 11.7 & 0.1 &  16.3\\
3351   & H\,{\sc ii} & 0.04 & 0.09&0.22 & 36.1 & 0.2  &  4.9\\
4303   & H\,{\sc ii} & 0.01 & 0.20& 0.28 & 23.8 & .2 & 14.4\\
4314   & LINER 2 & 0.05 & 0.11&0.43 & 8.6 & 0.3  &5.9\\
4321   & Transition 2 & 0.04 & 0.38& 0.22 & 15.8 & 0.2 & 21.7\\
5248   & H\,{\sc ii} & 0.03 &0.49& 0.10 & 4.6 & 0.5 &20.5\\
5953   & LINER 2, Sy2 & 0.13 &0.90 & 0.1 & 9.8 & 0.35 &7.7\\
6951   & Sy 2 & 0.03 &0.23 & 0.28 & 9.9 &0.25 &13.8\\
7742   & Transition 2 &0.18 &0.63 & 0.06 & 5.8 & 0.32 & 5.5\\
\hline
\end{tabular}
\end{center}
{\small Notes: Identification (col.~1); presence and type of nuclear activity (col.~2, from Ho et al. 1997b when available, NGC~1530 and NGC~5953 are from NED); relative nuclear ring size (col.~3) defined as ring semi-major axis (see Table~\ref{tab:phot}, col 11) divided by disk semi-major axis (half of the host major disk diameter), $D_0$ from NED (col~9); ring width (col.~4) defined as the difference between the isophotal inner ring boundary and outer ring boundary, divided by the disk semi-major axis; non-axisymmetric torque parameter (col.~5), from Com\'{e}ron \ea\ 2009; when no $Q_{b}$ value exists, we adopt a known $Q_{g}$ value instead: NGC~1530 \citep{B04} and NGC~5953 (H. Salo, private communication); rotation curve compactness (col.~6); rotation curve turnover radius (col.~7); We estimate that on average the uncertainty in the compactness is $\approx$75$\kms\kpc$. Uncertainties in the various size measurements involved were also added in quadrature to yield typical 
 uncertainties in the relative ring size and width  of $\pm0.01$. Uncertainties associated with the non-axisymmetric torque parameter, $Q_{\rm g}$, were adopted from Com\'{e}ron \ea\ (2009) to be $\pm0.1$.}
\end{table*}

\begin{figure*}
\center
\includegraphics[width=4.5in]{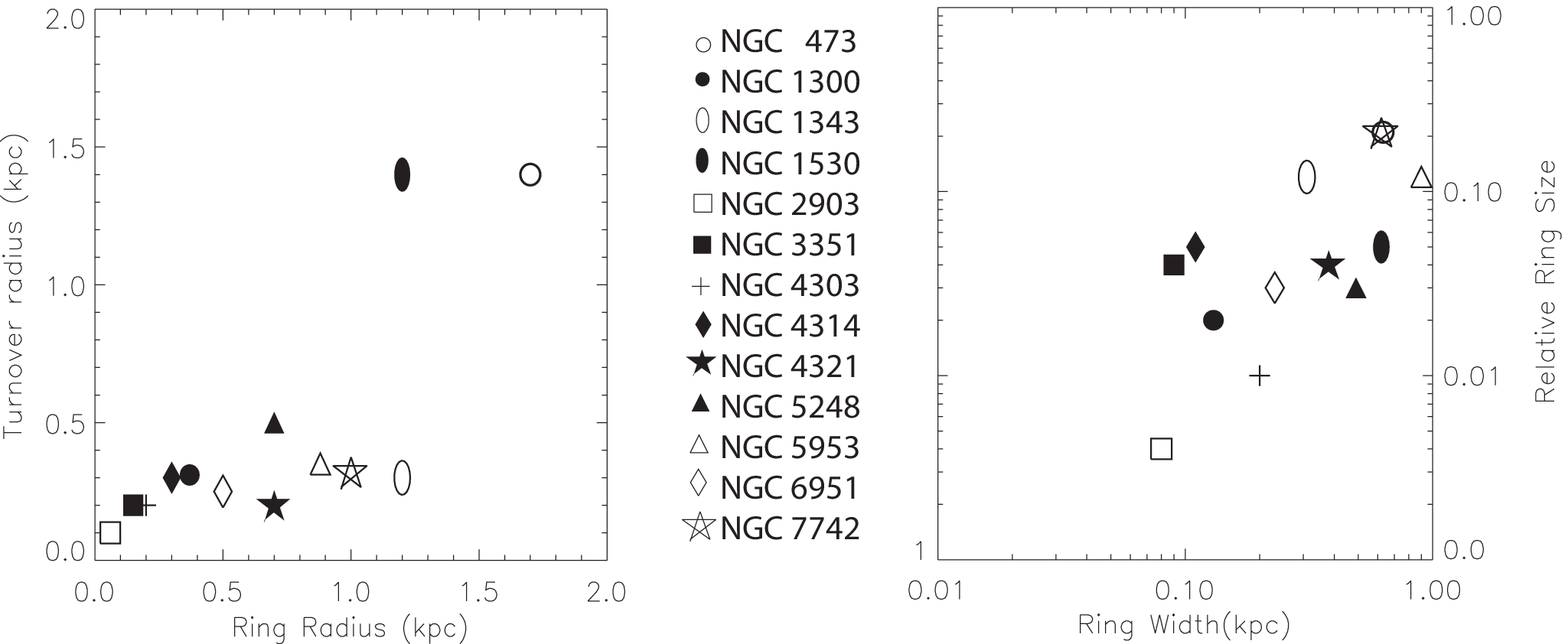}
\caption{Left panel: Radial location of the nuclear rings versus the rotation curve turnover radius. The locations generally coincide within the uncertainties. The spread is most likely due to the higher uncertainty associated with the ridge location (radial ring center) of the wider rings NGC~1343, NGC~4321, NGC~5953, and NGC~7742. Right panel: Ring width versus relative ring size. As the relative ring size increases so does the ring width.} 
\label{fig:turnover}
\end{figure*}

\begin{figure*}
\center
\includegraphics[width=6.5in]{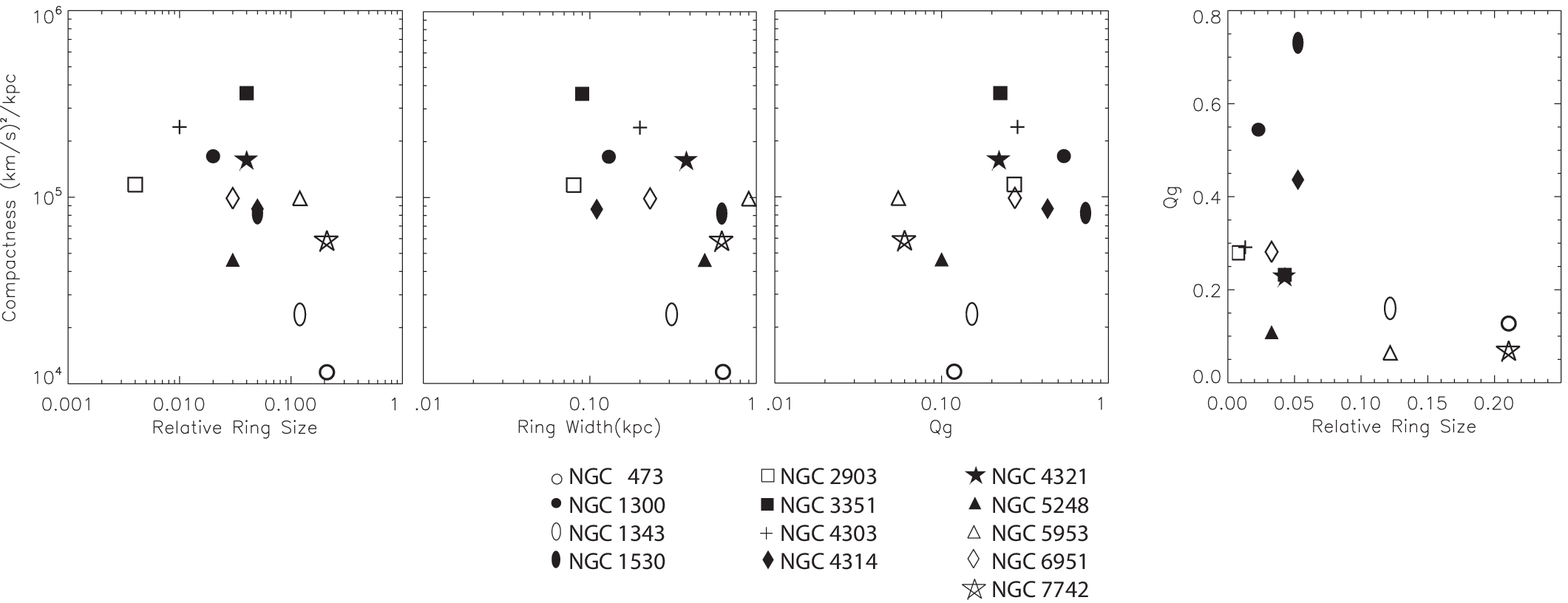}
\caption{Rotation curve compactness versus relative nuclear ring size (panel~1), nuclear ring width (panel~2), and the non-axisymmetric torque parameter $Q_{\rm g}$ (panel~3). Panel~4 shows the relative ring size as a function of  $Q_{\rm g}$. Typical error bars (refer to Table~\ref{tab:rc-parms} caption) are within the size of the icons.} 
\label{fig:rc-plots}
\end{figure*}

An indication of the central mass concentration can be obtained from the compactness, e.g., \cite{dB96}. We define compactness as $V^2/R_{\rm t}$, where $V$ is the rotational velocity difference between the radial location of the turnover radius, $R_{\rm t}$,  and the origin (0,0). The turnover radius is defined as the point in the rotation curve halfway between the initial steeply rising component and the flat(ter) segment. For the few galaxies with very steeply rising rotation curves, such as NGC~2903 and NGC~3351, the turnover radius cannot be measured accurately, and hence the compactness is uncertain. In these cases, the compactness should be  considered a lower limit. 

Table~\ref{tab:rc-parms} lists the compactness and turnover locations. As expected, the compactness decreases with increasing turnover radius (not shown graphically, but the values are in the Table. In linear theory, which is not strictly applicable here but yields useful approximations, the turnover radius in the rotation curve sets the location of the ILRs, and they in turn determine where inflowing gas piles up, which ultimately forms the nuclear ring \citep{JK95, BC96}. Indeed, the location of the nuclear rings correlates well with the turnover radius (see Figure~\ref{fig:rc} and the left panel of Figure~\ref{fig:turnover}), which we can identify with the phenomenon identified above.  

In panel~1 of Fig.~\ref{fig:rc-plots} we compare the relative ring size, that is the nuclear ring semi-major axis divided by the semi-major axis of the host galaxy, to the compactness parameter as derived from the rotation curve. We find that the relative ring size increases as the compactness decreases. 

The relative ring size is compared to the non-axisymmetric galactic torque, $Q_{\rm g}$, in panel four of Fig.~\ref{fig:rc-plots}. This torque parameter, originally defined by \cite{CS81} to characterize bar strength, quantifies the overall influence of galactic non-axisymmetries. Specifically, $Q_{\rm g}$ is the maximum tangential force at a given radius divided by the radial force. For all galaxies except two, we adopt the $Q_{\rm g}$ values from Comer\'{o}n et al. (2010), as noted in Table~\ref{tab:rc-parms}, since they applied a consistent computation method for all of the nuclear rings seen in our sample. We confirm (Fig.~\ref{fig:rc-plots}) that the highest torque values occur in those galaxies with the smaller rings relative to the size of their host disk (Knapen 2005; Comer\'{o}n et al. 2010). As the torque weakens below $Q_{\rm g}$ = 0.4 any size ring can form, but higher torque values ("stronger bars") progressively limit the nuclear ring size. 

The compactness and the non-axisymmetric torque $Q_{\rm g}$ do not correlate (panel 3 of Fig.~\ref{fig:rc-plots}). Also, the turnover and $Q_{\rm g}$ do not correlate (not shown graphically). This may be because the non-axisymmetric torque is a parameter driven more by the azimuthal than the radial distribution of mass, whereas the opposite is the case for the turnover radius and compactness. 

\subsection{Nuclear Ring Width Comparisons}

The width of a nuclear ring is an important parameter because it very much defines how well a ring is visible in an H$\alpha$ image, and how it is seen in contrast with the background emission from the host galaxy. Using the ring parameters we have now determined, it is interesting to see whether the ring width is driven by the physical properties of the host galaxy. So, as a first step, we define the ring width as the difference between the inner and outer ring boundaries, as defined from the emission in \halpha, and then plot the width of the nuclear ring against the relative ring size (right panel of Fig.~\ref{fig:turnover}). We see a trend that larger nuclear rings are wider, which might not have been expected {\it a priori}.

We then compare the compactness to the nuclear ring radial width (panel~2 of Fig.~\ref{fig:rc-plots}). There is some evidence for a trend of compactness with ring width, where a nuclear ring is wider as the compactness decreases. We can thus conclude that increased compactness leads to both smaller and narrower nuclear rings, which is in accord with the expectations from linear theory and the location of a pair of ILRs near the turnover radius of a rotation curve. This clear link between the kinematics of the galaxy and the morphological parameters of the nuclear rings is thus further direct evidence for a resonant origin of the nuclear rings, deeply linked to the underlying structure of the host galaxy. This evidence based on kinematics supplements earlier evidence based almost exclusively on morphology, as reviewed and extended by Comer\'on et al. (2010).

\section{Concluding Remarks}

We combined DensePak integral field unit and TAURUS Fabry-Perot observations for a sample of 13 galaxies that contain nuclear rings. Based on this sample, we note the following new findings:

\begin{itemize}
\item{Nuclear rings are intrinsically circular and unaffected by the local non-axisymmetric environment that can exist from active star formation within the rings; they are influenced by global phenomenon that can affect the potential, such as a bar.}
\item{The compactness, derived from the rotation curves, decreases as the ring width increases.}
\item{The compactness also decreases as the relative ring size increases}
\item{As the relative nuclear ring size increases so does the width of the ring}
\item{Strong bars with a $Q_{b}$ of 0.4 or greater can only be accompanied by small and thin nuclear rings. For smaller torques, the ring size can vary.}
\end{itemize}

We have also reaffirmed several standing findings:
\begin{itemize}
\item{True to resonance theory, the location of the nuclear rings correlates well with the turnover radius.} 
\item{The highest torque values occur in those galaxies with the smaller rings relative to the size of their host disk.}
\end{itemize}

In those cases with sufficient radial coverage, we agree that
\begin{itemize}
\item{For NGC~4314 residual velocities sharply decrease and plateau within the nuclear rings.  Ordered velocity patterns increase near the outer edge of the ring (both approaching and receding) close to both sides of the bar minor axis.}
\item{The residual velocities for NGC~5248 near the western side of the bar minor axis reach 40$\kms$, with a weaker-defined, although evident, transition zone as radii approach the outer ring boundary.}
\item{In the case of NGC~1530 large velocity excesses exist to the west of the nuclear ring, near its exterior edge and then sharply decrease over a small transition zone near the outer boundary of the nuclear ring.}
\end{itemize}

\acknowledgments

The Fabry-Perot observations were made with the William Herschel Telescope operated on the island of La Palma by the Newton Group of Telescopes in the Spanish Observatorio del Roque de los Muchachos of the Instituto de Astrof\'\i sica de Canarias.

The Integral Field Unit observations were made with the WIYN Observatory which is owned and operated by the WIYN Consortium, consisting of the University of Wisconsin, Indiana University, Yale University, and the National Optical Astronomy Observatory (NOAO). The WIYN is part of the Kitt Peak National Observatory located in Tucson Arizona.

\end{document}